\documentclass[12pt]{article}
\usepackage[dvips]{graphics}
\usepackage{graphicx}
\usepackage[body={5.9in, 9.0in},left=1.35in, top = 1.35in]{geometry}
\usepackage{setspace}
\newcommand{\x}{\textit{x} }
\newcommand{\D}{DGLAP equations }

\newcommand{\p}{\partial}
\newcommand{\be}{\begin{equation}}
\newcommand{\ee}{\end{equation}}
\newcommand{\bea}{\begin{eqnarray}}
\newcommand{\eea}{\end{eqnarray}}
\newcommand{\bes}{\begin{equation*}}
\newcommand{\ees}{\end{equation*}}
\newcommand{\ba}{\begin{array}}
\newcommand{\ea}{\end{array}}
\newcommand{\Fz}{ $ F_2^s(\frac{x}{z},t) $ }
\newcommand{\Gz}{ $ G(\frac{x}{z},t) $ }
\begin{document}
\title{Analytic and semi-analytic solution of the coupled DGLAP equations at small \textit{x} by the method of characteristics}
\author{D. K. Choudhury \thanks{e-mail: dkc\_phys@yahoo.co.in }\\Department of Physics, Gauhati University, Guwahati-781014, India \and
  P. K. Sahariah\thanks{e-mail: pksahariah@yahoo.com }\\Department of Physics, Cotton College, Guwahati-781001, India }
\date{}

\maketitle
\begin{abstract}

Coupled DGLAP equations involving singlet quark and gluon distributions are explored by a Taylor expansion at small \textit{x} as two first order partial differential equations in two variables : Bjorken \textit{x} and \textit{t} ($t =log\frac{Q^2}{\Lambda^2}$).The system of equations are then reduced to canonical form and the resultant equations are solved by the method of characteristics.Analytic and semi-analytic solutions thus obtained are compared with the exact results and the range of validity obtained.

\end{abstract}
\section{Introduction}
\label{ch5intro}
DGLAP equations \cite{AP,GRIBOV,LIPATOV,DOKSH} have been playing very important role in the study of nucleon structure function in deep inelastic regime.In the standard procedure, the $x$ profile of the parton distribution at some low scales $Q^2_0$ are parametrized by comparing with data while their values at a higher scale are obtained by evolution of these equations, which are solved numerically.
In the small  \textit{x} regime however, several analytical solutions\cite{Ch2BallFortePLB335,Ch2BallFortePLB336,KotikovNPB549,GehrmannPLB365,ManSWPLB393} of the DGLAP equations are available in the literature, which are in good agreement with HERA data \cite{Cooper,H1EPJC13}. This suggests utility of such approaches in understanding the dynamics of evolution of quarks and gluons at small  \textit{x}.
In recent years the present authors and their collaborators \cite{RDDKCZPhyC75,JSDKCGMPLB403,DKCRDASEPJC2,PKSPram58} have also been pursuing an approximate method of solution of DGLAP equations at small \x with considerable phenomenological success

 In this paper we present solutions of the coupled singlet evolution equations applying the method of characteristics. By a Taylor series expansion of the gluon and the singlet distribution that occur under the integral sign of  the coupled equations, we convert the coupled integro-differential equations  into two  first order partial differential equations in two variables : Bjorken \textit{x} and $ t (t=ln(Q^2/\Lambda^2))$. The resulting system of  equations which are still coupled, is then reduced to canonical form by the introduction of a suitable new function. After reduction to  the canonical forms, the equations can be solved by applying the method of characteristics\cite{Farlow,LEWilliams,Zachmanoglou} under certain approximations valid to be at small \textit{x}. 

In \S\ref{formalism}, we describe the formalism, \S\ref{results} is devoted to testing our prediction with the exact solution\cite{MRST2001} and  data\cite{H1EPJC21,H1EPJC13} ,while in \S\ref{con:ch5conclusion} we give our conclusion.

\section{Formalism \label{formalism}}
\subsection{Singlet coupled DGLAP equations in Taylor approximated form}The coupled DGLAP equations  for quark singlet $(\Sigma(x,Q^2))$ and gluon $(g(x,Q^2))$ densities are \cite{AP,GRIBOV,LIPATOV,DOKSH}
\be
\label{eqn:ch5singcoup}
\frac{\p}{\p logQ^2}\left(\ba{c}
\Sigma\left(x,Q^2\right)\\
g\left(x,Q^2\right)
\ea
\right) = \frac{\alpha_s\left(Q^2\right)}{2\pi}\left(
\ba{cc}
P_{qq} & P_{qg} \\
P_{gq} & P_{gg}
\ea
\right)\otimes \left(
\ba{c}
\Sigma\left(x,Q^2\right) \\
g\left(x,Q^2\right)
\ea
\right) ,
\ee
where $\alpha_s\left(Q^2\right)$ is the strong coupling constant, $ P_{i,j}$s are the Altarelli-Parisi splitting functions and the symbol $\otimes$ stands for the usual Mellin convolution in the first variable already defined as
\be
\label{eqn:ch5Mellin}
 a(x)\otimes f(x)=\int_x^1\,\frac{dy}{y}a(y)f\left(\frac{x}{y}\right).
\ee
Introducing the variable $ t=ln\frac{Q^2}{\Lambda^2}$ and using the explicit forms of the splitting functions  $ P_{i,j}$ in LO,  Eq.(\ref{eqn:ch5singcoup}) can be written as \cite{LFAbbot}
\begin{eqnarray}
\label{eqn:ch5F2s1}
&\displaystyle{\frac{\p F_2^S(x,t)}{\p t}}-\frac{A_f}{t}\left[\{3+4ln(1-x)\}F_2^S(x,t)+2\int_x^1\frac{dz}{(1-z)}\left\{(1+z^2)F_2^s\left(\frac{x}{z},t\right)\right.\right. \nonumber \\
& \left.\left.-2F_2^S(x,t)\right\}+\frac{3}{2} n_f \int_x^1dz\left(z^2+(1-z)^2\right)G\left(\frac{x}{z},t\right) \right] =  0
\end{eqnarray}
and
\begin{eqnarray}
\label{eqn:ch5Gluon1}
\frac{\p G(x,t)}{\p t} - \frac{A_f}{t}\left[\left\{\frac{11}{12} - \frac{n_f}{18} + ln (1-x)\right\}G(x,t)\right.& & \nonumber \\
\left.+ \int_x^1 dz\left\{ \frac{z\, G\left(\frac{x}{z},t\right) - G(x,t)}{1-z}+ \left(z\,(1-z) + \frac{1-z}{z}\right)G\left(\frac{x}{z},t\right)\right\}\right.\nonumber & & \\
\left.+\frac{2}{3}\int_x^1dz\left(\frac{1+(1-z)^2}{z}\right)F_2^s\left(\frac{x}{z},t\right)\right]& = & 0 .
\end{eqnarray}
Here $A_f=\frac{4}{3\beta_0}$, $ \beta_0=11-\frac{2}{3}n_f $  and  
$\alpha_s(t)=\frac{4\pi}{\beta_0 t}$ .  $ G(x,t)(=xg(x,t))$ is the gluon momentum distribution and  $  F_2^S(x,t)$ is the singlet structure function of the proton defined as
\be
\label{eqn:ch5singsf}
F_2^S(x,t)=x\Sigma(x,t)=x\sum_{i=1}^{n_f}\left (q_i(x,t)+\overline{q}_i(x,t)\right),
\ee
where $q_i(x,t)$ is the quark distribution of the $i$ th flavour inside the proton.
We write Eqs.(\ref{eqn:ch5F2s1}) and (\ref{eqn:ch5Gluon1}) as
\begin{eqnarray}
\label{eqn:ch5F2s2}
\displaystyle{\frac{\p F_2^S(x,t)}{\p t}}-\frac{A_f}{t}\left[\{3+4ln(1-x)\}F_2^S(x,t)+I_1^s(x,t)+I_1^g(x,t)\right]=0
\end{eqnarray}
and
\begin{eqnarray}
\label{eqn:ch5Gluon2}
\displaystyle{\frac{\p G(x,t)}{\p t}}-\frac{9A_f}{t}\left[\left\{\frac{11}{12}-\frac{n_f}{18}+ln(1-x)\right\}G(x,t)+I_2^s(x,t)+I_2^g(x,t)\right]=0 ,
\end{eqnarray}
where
\be
\label{eqn:ch5I1s}
I_1^s(x,t)=2\int_x^1\frac{dz}{1-z}\left[(1+z^2)F_2^S(\frac{x}{z},t)-2F_2^S(x,t)\right] ,
\ee
\be
\label{eqn:ch5I1g}
I_1^g(x,t)=\frac{3}{2} n_f \int_x^1 dz\,[z^2+(1-z)^2]G(\frac{x}{z},t) ,
\ee
\be
\label{eqn:ch5I2s}
I_2^s(x,t)=\frac{2}{3}\int_x^1dz\left[\left(\frac{1+(1-z)^2}{z}\right)F_2^S\left(\frac{x}{z},t\right)\right]
\ee
and
\be
\label{eqn:ch5I2g}
I_2^g(x,t)= \int_x^1 dz\left\{ \frac{z\,G\left(\frac{x}{z},t\right) - G(x,t)}{1-z}+ \left(z\,(1-z) + \frac{1-z}{z}\right)G\left(\frac{x}{z},t\right)\right\}.
\ee
To carry out the integrations in Eqs.(\ref{eqn:ch5I1s}-\ref{eqn:ch5I2s}), we introduce the variable $ u$  defined as $u=1-z $
 and  expand the argument $\displaystyle{\frac{x}{z}}$ in \Fz  and \Gz  as a series.
\be
\label{eqn:ch5xbyz}
\frac{x}{z}=\frac{x}{1-u}=x\sum_{k=0}^\infty u^k=x+x\sum_{k=1}^\infty u^k .
\ee
Using Eq.( \ref{eqn:ch5xbyz})  we expand \Fz and \Gz in Taylor series as:
\be
\label{eqn:ch5F2series}
F_2^S(\frac{x}{z},t)=F_2^S(x,t)+x\sum_{k=1}^\infty u^k \frac{\p F_2^S(x,t)}{\p x}+\frac{(
x\sum_{k=1}^\infty u^k)^2}{2!} \frac{\p^2 F_2^S(x,t)}{\p x^2} + ..
\ee
and
\be
\label{eqn:ch5Gseries}
G(\frac{x}{z},t)=G(x,t)+x\sum_{k=1}^\infty u^k \frac{\p G(x,t)}{\p x}+\frac{(
x\sum_{k=1}^\infty u^k)^2}{2!} \frac{\p^2 G(x,t)}{\p x^2} + ..
\ee
The series Eqs. (\ref{eqn:ch5F2series}) and (\ref{eqn:ch5Gseries}) are convergent \cite{JSDKCGMPLB403,PKSPram58} and hence at small \x , we can approximate these by
\be
\label{eqn:ch5F2seriesApp}
F_2^S(\frac{x}{z},t)\approx F_2^S(x,t)+x\sum_{k=1}^\infty u^k \frac{\p F_2^S(x,t)}{\p x}
\ee
and
\be
\label{eqn:ch5GseriesApp}
G(\frac{x}{z},t)\approx G(x,t)+x\sum_{k=1}^\infty u^k \frac{\p G(x,t)}{\p x}.
\ee
Using the above two  Eqs.(\ref{eqn:ch5F2seriesApp}) and (\ref{eqn:ch5GseriesApp})  we carry out the integrations in $z$ in Eqs.(\ref{eqn:ch5I1s}-\ref{eqn:ch5I2s}). Neglecting terms $O(x^2)$ which is justified at small \x, we get
\be
\label{eqn:ch5I1sAp}
I_1^s(x,t)\approx(2x-3)F_2^S(x,t)+\left(x+2 x \ln\frac{1}{x}\right)\frac{\p F_2^s(x,t)}{\p x},
\ee
\be
\label{eqn:ch5I1gAp}
I_1^g(x,t)\approx n_f\left(1-\frac{3}{2}x\right)G(x,t)-\frac{n_f}{2}\left(5x-3 x \ln\frac{1}{x}\right)\frac{\p G(x,t)}{\p x},
\ee
\be
\label{eqn:ch5I2sAp}
I_2^s(x,t)\approx \left(-x+\frac{4}{3}x\ln\frac{1}{x}\right)F_2^S(x,t)+\frac{4}{3} x \frac{\p F_2^S(x,t)}{\p x} ,
\ee
and
\be
\label{eqn:ch5I2gAp}
I_2^g(x,t)\approx R'_g(x)G(x,t)+P_g(x)\frac{\p G(x,t)}{\p x},
\ee
where the function $P_g(x)$ is 
\be
P_g(x)=x(\ln\frac{1}{x}-\frac{11}{12}) ,
\ee
and  $R'_g(x)$  is given by
\be
 R'_g(x)=(ln\frac{1}{x}-\frac{11}{6})+2x .
\ee
Using Eqs.( \ref{eqn:ch5I1sAp}-\ref{eqn:ch5I2gAp}), we recast Eq.(\ref{eqn:ch5F2s2}) and Eq.(\ref{eqn:ch5Gluon2}) as two first order partial differential equations in \x and \textit{t} in standard form:
\begin{eqnarray}
\label{eqn:ch5pde1}
 a_{11}\frac{\p F_2^S(x,t)}{\p t}+a_{12}\frac{\p G(x,t)}{\p t}+b_{11}\frac{\p F_2^S(x,t)}{\p x}+b_{12}\frac{\p G(x,t)}{\p x}\nonumber \\
=R_{11}F_2^S(x,t)+R_{12}G(x,t)
\end{eqnarray}
and
\begin{eqnarray}
\label{eqn:ch5pde2}
 a_{21}\frac{\p F_2^S(x,t)}{\p t}+a_{22}\frac{\p G(x,t)}{\p t}+b_{21}\frac{\p F_2^S(x,t)}{\p x}+b_{22}\frac{\p G(x,t)}{\p x} \nonumber \\
=R_{21}F_2^S(x,t)+R_{22}G(x,t) ,
\end{eqnarray}
where
\begin{eqnarray}
\label{eqn:ch5aelements}
\left. \begin{array}{ll}
a_{11}=t , & a_{12}=0 \\
a_{21}=0 , & a_{22}=t
\end{array}\right\},
\end{eqnarray}
\begin{eqnarray}
\label{eqn:ch5belements}
\left. \begin{array}{ll}
b_{11}=-12 A_fx , \hspace{0.3in}  &  b_{12}=\frac{A_fn_f}{2}\left(5x-3x\ln\frac{1}{x}\right) \\
b_{21}=-12 A_fx , &  b_{22}=-9A_f P_g(x)
\end{array}\right\}
\end{eqnarray}
and
\begin{eqnarray}
\label{eqn:ch5Relements}
\left. \begin{array}{ll}
R_{11}=A_f(2x+4\ln(1-x)), &  R_{12}=A_f n_f(1-\frac{3}{2}x)  \\
R_{21}=9A_f\left (-x+\frac{4}{3}x\ln\frac{1}{x}\right),&  R_{22}=9A_f\left(\frac{11}{12}-\frac{n_f}{18}+\ln(1-x)+R'_g(x)\right)
\end{array} \right\}.
\end{eqnarray}
Eq.(\ref{eqn:ch5pde1}) and Eq.\ref{eqn:ch5pde2}) are two first order linear coupled differential equations. We now reduce these two equations to canonical form where they are decoupled from each other in terms of a new function. To that end we introduce a vector  $ \vec{u}(x,t) $  as 
\be
\label{eqn:ch5vecu}
\vec{u}(x,t)=\left(\ba{c}

F_2^S(x,t)  \\
G(x,t)
\ea
\right)
\ee
and write the two equations Eqs.(\ref{eqn:ch5pde1}) and (\ref{eqn:ch5pde2}) in matrix form:
\be
\label{eqn:ch5pdemf}
a\, \vec{u}_t(x,t)+b\,\vec{u}_x(x,t)=R\,\vec{u}(x,t),
\ee
where the matrices $a$, $b$ and $R$ are given by
\be
\label{eqn:ch5mata}
a=\left(\ba{cc}
a_{11} & a_{12} \\
a_{21} & a_{22}
\ea
\right) ,
\ee
\be
\label{eqn:ch5matb}
b=\left(\ba{cc}
b_{11} & b_{12} \\
b_{21} & b_{22}
\ea
\right)
\ee
and
\be
\label{eqn:ch5matR}
R=\left(\ba{cc}
R_{11} & R_{12} \\
R_{21}& R_{22}
\ea
\right)
\ee
In Eq.(\ref{eqn:ch5pdemf}), $u_t$ and $u_x$ are the derivatives with respect to $t$ and $x$ respectively. We note that the matrix  $a$ with its elements given by Eq.(\ref{eqn:ch5aelements}) is a non-singular matrix and hence multiplying Eq.(\ref{eqn:ch5pdemf}) by $a^{-1}$ from left, we get 
\be
\label{eqn:ch5pdeAB}
  \vec{u}_t(x,t)+A\,\vec{u}_x(x,t)=B\,\vec{u}(x,t),
\ee
where the new matrices $A$ and $B$ are:
\be
\label{eqn:ch5A}
A=a^{-1}b
\ee
and
\be
\label{eqn:ch5B}
B=a^{-1}R .
\ee
Equation (\ref{eqn:ch5pdeAB}) is a system of two coupled first order linear partial differential equations in the two variables \x and \textit{t}  for the vector $  \vec{u} $ prescribed by Eq.(\ref{eqn:ch5vecu}). Its principal part, i.e. $ \vec{u}_t(x,t)+A\vec{u}_x(x,t) $ is completely characterized by the coefficient matrix A. Since the matrix $ A$ has n [here n=2] distinct eigenvalues, the system Eq.(\ref{eqn:ch5pdeAB} ) is a hyperbolic one and it is possible to obtain its canonical form in the following way \cite{LEWilliams,Zachmanoglou}:

Let $\lambda_1 , \lambda_2 $ be the two distinct and real eigenvalues of the matrix $ A $ and let $ \vec{p}_1$ and $ \vec{p}_2 $ be the corresponding eigenvectors. Let $P$ be a 2x2 matrix  formed by the eigenvectors $ \vec{p}_1$, $ \vec{p}_2 $, i.e.
\be
\label{eqn:ch5matrixP}
P=(\ba{cc} \vec{p_1} \hspace{0.3in} \vec{p_2} ) .
\ea
\ee
Now if $ \Lambda $ is the diagonal matrix with the eigenvalues $\lambda_1$ and $ \lambda_2$ as the two elements, then we have
\be
\label{eqn:ch5PAP}
P^{-1}AP\equiv\Lambda=\left(\ba{cc}
\lambda_1 & 0 \\
0 & \lambda_2 
\ea \right).
\ee
Let us define a new vector $ \vec{v} $ by the relation
\be
\label{eqn:ch5pinvu}
\vec{v}=P^{-1}\vec{u}
\ee
so that 
\be
\label{eqn:ch5upv}
\vec{u}=P\vec{v}.
\ee
For convenience, we have dropped the functional form in $\vec {u}(x,t)$, $\vec{v}(x,t)$ and $P(x,t)$. Differentiating Eq.(\ref{eqn:ch5upv}) with respect to $ t$ and $x$ respectively we get
\be
\label{eqn:ch5utux}
\vec{u}_t=P\vec{v}_t+P_t \vec{v} , \hspace{0.4in}\vec{u}_x=P\vec{v}_x+P_x \vec{v}.
\ee
Substituting Eqs.(\ref{eqn:ch5upv}) and (\ref{eqn:ch5utux}) in Eq.(\ref{eqn:ch5pdemf}), we obtain
\be
\label{eqn:ch5ptpx}
P\vec{v}_t+P_t \vec{v}+A P\vec{v}_x+A P_x \vec{v}=BP\vec{v}.
\ee
Multiplying Eq.(\ref{eqn:ch5ptpx}) from left by $ P^{-1}$ and using Eq.(\ref{eqn:ch5PAP}), we obtain 
\be
\label{eqn:ch5vtvx}
\vec{v}_t+\Lambda \vec{v}_x=\vec{e},
\ee
where
\be
\label{eqn:ch5vece}
\vec{e}=P^{-1}(BP-P_t-AP_x) \vec{v}
\ee
is a two component column matrix.
Eq.(\ref{eqn:ch5vtvx}) is in canonical form. In component form it is 
\be
\label{eqn:ch5comf}
\frac{\p v_i}{\p t}+\lambda_i \frac{\p v_i}{\p x}=e_i. \hspace{0.7in}i=1,2
\ee
\subsection{Solution by the method of characteristics}
\label{subs:ch5solbymethodofch}
Equation (\ref{eqn:ch5comf}) shows that the principal part of the $ i$ th equation, viz.  {$\displaystyle{\frac{\p v_i}{\p t}+\lambda_i \frac{\p v_i}{\p x}}$} involves only the component $v_i$ of the vector $\vec v$ and its derivatives. To solve these equations by the method of characteristics, we define first the characteristic curve of the system .These are the curves in the $  x-t $ plane given by $ x=x_i(t) $, where $x_i(t) $ is a solution of the differential equation
\be
\label{eqn:ch5cheq}
\frac{d x_i(t)}{d t}=\lambda_i(x,t)
\ee
with  $ \lambda_i(x,t)$ being the eigenvalues of the coefficient matrix $ A(x,t) $ defined in Eq.(\ref{eqn:ch5A}). It is also known \cite{Zachmanoglou} that the characteristic curves of a hyperbolic system like Eq.(\ref{eqn:ch5pdeAB}) whose eigenvalues $ \lambda_i(i=1,2)$ are distinct, remain invariant under transformation of the system to its canonical form Eq.(\ref{eqn:ch5comf}). We also make the observation that along a characteristic curve corresponding to an eigenvalue $ \lambda_i$, the left hand side ( i.e. the principal part) of Eq.(\ref{eqn:ch5comf}) is actually an ordinary derivative with respect to $ t$, since 
\be
\frac{d v_i(t)}{d t}=\frac{\p v_i(x_i(t),t)}{\p t}+\frac{\p v_i(x_i(t),t)}{\p x}\frac{d x_i(t)}{d t},
\ee
where $\displaystyle  \frac{d x_i(t)}{d t}$ is given by Eq.(\ref{eqn:ch5cheq}) on the characteristic curve $ x=x_i(t)$. As a result, Eq.(\ref{eqn:ch5comf}) becomes an ordinary differential equation:
\be
\label{eqn:ch5odevt}
\frac{d v_i(t)}{d t}=e_i\left( x_i(t),t,\vec{v}(x_i(t),t)\right)
\ee
along the characteristic curves.
The actual integration on the right hand side of Eq.(\ref{eqn:ch5odevt}) depends on the analytical solution of Eq.(\ref{eqn:ch5cheq}). It, in turn depends on the eigenvalues $\displaystyle \lambda_1$ and $\displaystyle \lambda_2$ of the coefficient matrix $ A(x,t)$ of Eq.(\ref{eqn:ch5pdeAB}). On calculating these eigenvalues and the eigenvectors with matrix $\displaystyle a$ and $b$ defined in Eqs(\ref{eqn:ch5mata}) and (\ref{eqn:ch5matb}), we find that they are too involved as discussed briefly in Appendix. As a result, the characteristic equation (\ref{eqn:ch5cheq}) cannot be solved analytically to find the curve $  x=x_i(t)$.

The situation can be simplified by making some further approximations in Eqs.(\ref{eqn:ch5I1sAp}- \ref{eqn:ch5I2gAp}). If we keep only leading terms $\displaystyle  x\,\ln\frac{1}{x}$ compared to terms $ O(x)$, then Eqs.(\ref{eqn:ch5I1sAp}-\ref{eqn:ch5I2gAp}) become respectively
\be
\label{eqn:ch5I1sApf}
I_1^s(x,t)\approx -3 F_2^S(x,t)+\left(2\, x \ln\frac{1}{x}\right)\frac{\p F_2^S(x,t)}{\p x},
\ee
\be
\label{eqn:ch5I1gApf}
I_1^g(x,t)\approx n_f G(x,t)-\frac{n_f}{2}\left(-3 x\,\ln\frac{1}{x}\right)\frac{\p G(x,t)}{\p x},
\ee
\be
\label{eqn:ch5I2sApf}
I_2^s(x,t)\approx \left(\frac{4}{3} x\, ln\frac{1}{x}\right)F_2^S(x,t)
\ee
and
\be
\label{eqn:ch5I2gApf}
I_2^g(x,t)\approx R^\prime_g(x)G(x,t)+P_g(x)\frac{\p G(x,t)}{\p x}
\ee
with
\be
 R^\prime_g(x)=\left(\ln\frac{1}{x}-\frac{11}{6}\right)
\ee
and
\be
\label{eqn:ch5pgx}
P_g(x)=x\,\ln\frac{1}{x}.
\ee
The matrices $b$ (Eq.\ref{eqn:ch5matb}) and $R$ (Eq.\ref{eqn:ch5matR}) are now given by
\be
\label{eqn:ch5b}
b=\left(\ba{cc}
-2A_f x\,\ln\frac{1}{x} & -\frac{3}{2}A_f n_f x\,\ln\frac{1}{x} \\
0 & \hspace{0.3in} -9A_f P_g(x)
\ea
\right)
\ee
and
\be
\label{eqn:ch5R}
R=\left(\ba{cc}
0 & A_f n_f \\
12 A_f x\,\ln\frac{1}{x} & \hspace{0.3in}  9A_f\left(\frac{11}{12}-\frac{n_f}{18}+R^\prime_g(x)\right)
\ea
\right).
\ee
The eigenvalues of the matrix $ A $ are then obtained from the characteristic equation
\be
\label{eqn:ch5detA}
det\mid A-\lambda I\mid=0
\ee
leading to 
\be
\label{eqn:ch5lambda1}
\lambda_1=-\frac{9A_f x\,ln\frac{1}{x}}{t}
\ee
and
\be
\label{eqn:ch5lambda2}
\lambda_2=-\frac{2A_f x\,\ln\frac{1}{x}}{t}.
\ee
The corresponding eigenvector matrix $ P $ (Eq.\ref{eqn:ch5matrixP}) is 
\be
\label{eqn:ch5P}
P=\left(\ba{cc}
3n_f & 1 \\
1 & 0
\ea\right).
\ee
This simplification yields the column matrix $ \vec{e}=\left(\ba{c} e_1 \\ e_2  \ea \right)$ defined in Eq.(\ref{eqn:ch5vece}) and occurred in Eq.(\ref{eqn:ch5comf}) with the components
\be
\label{eqn:ch5compe1}
e_1=\left(\frac{12A_f x\,\ln\frac{1}{x}}{t}\right)v_2+\left(\frac{18 A_f n_f x\,\ln\frac{1}{x}}{7t}+\frac{9A_f(-\frac{11}{12}-\frac{n_f}{18}+ln\frac{1}{x})}{t}\right)v_1 \hspace{1.0in}
\ee
and
\be
\label{eqn:ch5compe2}
e_2=\frac{-18 A_f n_fx\,\ln\frac{1}{x}}{7t}v_2+\left(\frac{A_f n_f}{t}-\frac{27 A_f n_f^2 x\,\ln\frac{1}{x}}{49t}-\frac{27A_fn_f}{14t}\left(-\frac{11}{12}-\frac{n_f}{18}+ln\frac{1}{x}\right) \right)v_1 .
\ee
Eqs.(\ref{eqn:ch5compe1}) and (\ref{eqn:ch5compe2}) will be used to solve Eq.(\ref{eqn:ch5odevt}) for $ v_1(t)$ and $ v_2(t)$ on the characteristic curves corresponding to eigenvalues $\lambda_1$ and $ \lambda_2 $.

Let $ (\bar{x},\bar{t})$ be a fixed point in the $(x-t)$ plane through which the two characteristic curves $x_1(t)$ and  $ x_2(t)$ will pass i.e.
\be
\label{eqn:ch5x1x2}
  x_1(\bar{t})=\bar{x} \hspace{0.3in} and\hspace{0.3in}   x_2(\bar{t})=\bar{x} .
\ee
 Explicitly, the equations of the two characteristics (Eqs.(\ref{eqn:ch5cheq})) are 
\be
\label{eqn:ch5cheqx1}
\frac{d x_1(t)}{d t}=-\frac{9 A_f x\,\ln\frac{1}{x}}{t}
\ee
and
\be
\label{eqn:ch5cheqx2}
\frac{d x_2(t)}{d t}=-\frac{2 A_f x\,\ln\frac{1}{x}}{t}
\ee
satisfying the condition of passing through the point $(\bar{x},\bar{t}) $ as given by Eq.(\ref{eqn:ch5x1x2}). Solution of Eq.(\ref{eqn:ch5cheqx1}) is
\be
\label{eqn:ch5chcur1}
\ln\,x_1(t)=\ln\,\bar{x}\left(\frac{t}{\bar{t}}\right)^{9A_f} ,
\ee
while that of Eq.(\ref{eqn:ch5cheqx2}) is
\be
\label{eqn:ch5chcur2}
\ln\,x_2(t)=\ln\bar{x}\left(\frac{t}{\bar{t}}\right)^{2A_f} .
\ee
Furthermore, if the characteristic curves cut the initial curve $ t=t_0(=\ln\frac{Q_0^2}{\Lambda^2}) $ at $ x_1=\tau_1 $ and $ x_2=\tau_2$ respectively [Fig.1], then Eqs.(\ref{eqn:ch5chcur1}) and (\ref{eqn:ch5chcur2}) give
\be
\label{eqn:ch5lntau1}
\ln\tau_1=\ln\bar{x}\left(\frac{t_0}{\bar{t}}\right)^{9A_f}
\ee
and
\be
\label{eqn:ch5lntau2}
\ln\tau_2=\ln\bar{x}\left(\frac{t_0}{\bar{t}}\right)^{2A_f}
\ee
leading to 
\be
\label{eqn:ch5tau1}
\tau_1=\bar{x}^{(\frac{t_0}{\bar{t}})^{9A_f}}
\ee
and
\be
\label{eqn:ch5tau2}
\tau_2=\bar{x}^{(\frac{t_0}{\bar{t}})^{2A_f}} .
\ee
Dropping the bars over $ \bar{x} $ and $ \bar{t}$ , following are the expressions for the characteristics 
\be
\label{eqn:ch5x1t}
x_1(t)=\tau_1^{\left(\frac{t_0}{t}\right)^{-9A_f}}
\ee
and
\be
\label{eqn:ch5x2t}
x_2(t)=\tau_2^{\left(\frac{t_0}{t}\right)^{-2A_f}} .
\ee
Let us now find the solution of Eq.( \ref{eqn:ch5odevt}) which has explicit forms on the characteristic curves as 
\be
\label{eqn:ch5odev1}
\frac{d v_1(t)}{v_1(t)}=c_1\left(x_1(t),t,v_1(x_1(t),t),v_2(x_1(t),t)\right)dt
\ee
and
\be
\label{eqn:ch5odev2}
\frac{d v_2(t)}{v_1(t)}=c_2\left(x_2(t),t,v_1(x_2(t),t),v_2(x_2(t),t)\right)dt,
\ee
where
\begin{eqnarray}
\label{eqn:ch5constc1}
c_1\left(x_1(t),t,v_1(x_1(t),t),v_2(x_1(t),t)\right)=\displaystyle \frac{12 A_f x_1(t)}{t} ln\frac{1}{x_1(t)} \left( \frac{v_2(x_1(t),t)}{v_1(x_1(t),t)}\right) & & \nonumber \\
+ \left\{ \frac{18 A_f n_f}{7} x_1(t)ln\frac{1}{x_1(t)}+9 A_f \left(-\frac{11}{12}-\frac{n_f}{18}+ln\frac{1}{x_1(t)}\right)\right\} \frac{1}{t}
\end{eqnarray}
and
\begin{eqnarray}
\label{eqn:ch5constc2}
c_2\left(x_2(t),t,v_1(x_2(t),t),v_2(x_2(t),t)\right)=-\frac{18 A_f n_f x_2(t) ln\frac{1}{x_2(t)}}{t}+ 
\left\{A_f n_f-\right. & & \nonumber \\
\left.\frac{27}{49} A_f n_f^2 x_2(t) ln\frac{1}{x_2(t)}+\frac{27}{14} A_f n_f \left(\frac{11}{12}+\frac{n_f}{18}\right)-\frac{27}{14}A_f n_f ln\frac{1}{x_2(t)}\right\} \frac{1}{t}\left(\frac{v_1(x_2(t),t)}{v_2(x_2(t),t)}\right). 
\end{eqnarray}
Integration of Eqs.(\ref{eqn:ch5odev1}) and (\ref{eqn:ch5odev2}) along the characteristic curves $ x_1(t)$ and $ x_2(t)$ from $t_0$ and $ \bar{t}$  gives:
\be
\label{eqn:ch5v1xtbar}
v_1(\bar{x},\bar{t})=v_1(\tau_1)exp\,\left[\int_{t_0}^{\bar{t}}c_1(x_1(t),t,v_1(x_1(t),t),v_2(x_1(t))dt\right] \, ,
\ee
\be
\label{eqn:ch5v2xtbar}
v_2(\bar{x},\bar{t})=v_2(\tau_1)exp\,\left[\int_{t_0}^{\bar{t}}c_1(x_2(t),t,v_1(x_2(t),t),v_2(x_2(t))dt\right].
\ee
In Eqs.(\ref{eqn:ch5v1xtbar}) and (\ref{eqn:ch5v2xtbar}), we have now the boundary conditions $v_1(\tau_1) $ and $v_2(\tau_1) $ at $ x=\tau_1 $  and $ x=\tau_2$   where the two characteristic curves meet the initial line  $ t=t_0$(Fig1).
Eqs.(\ref{eqn:ch5v1xtbar}) and (\ref{eqn:ch5v2xtbar}) give the two components $v_1$ and $v_2$ of the function $\vec{v}$ defined in Eq.(\ref{eqn:ch5pinvu}). Now, from Eq.(\ref{eqn:ch5pinvu}) with $P^{-1}$ found from Eq.(\ref{eqn:ch5P}), we find that the components $v_1$ and $v_2$ are related to singlet structure function $F_2^S(x,t)$ and gluon momentum distribution $G(x,t)$ by
\be
\label{eqn:ch5v1xt}
v_1(x,t)=G(x,t),
\ee
and
\be
\label{eqn:ch5v2xt}
v_2(x,t)=F_2^S(x,t)-\frac{3 n_f}{14} G(x,t).
\ee
We can therefore rewrite Eqs.(\ref{eqn:ch5v1xtbar}) and (\ref{eqn:ch5v2xtbar}) as
\be
\label{eqn:ch5Gxt1}
G(x,t)=G(\tau_1)\exp\left[\int_{t_0}^t c_1(x_1(t'),t',v_1(x_1(t'),t'),v_2(x_1(t'),t'))dt'\right]
\ee
and
\bea
\label{eqn:ch5F2sxt}
F_2^S(x,t)=&\frac{3 n_f}{14} G(x,t)+\left(F_2^S(\tau_2)-\frac{3 n_f}{14}G(\tau_2)\right) \nonumber \\
&\exp\left[\int_{t_0}^{t} c_2\left(x_2(t'),t',v_1(x_2(t'),t'),v_2(x_2(t'),t')\right)dt'\right].
\eea
In writing Eqs.(\ref{eqn:ch5Gxt1}) and (\ref{eqn:ch5F2sxt}) we have removed the bar over $ \bar{x}$ and $ \bar{t}$ and written the integration variable as $ t' $. Eq.(\ref{eqn:ch5Gxt1}) and Eq.(\ref{eqn:ch5F2sxt}) are the most general solutions for the gluon and the singlet distribution within the present formalism.

Unfortunately, the right hand side of both Eq. (\ref{eqn:ch5Gxt}) and Eq.(\ref{eqn:ch5F2sxt})[see also Eqs.(\ref{eqn:ch5constc1}) and (\ref{eqn:ch5constc2})] contain the ratio $\displaystyle  \left( \frac{v_2}{v_1}\right)$ which is yet unknown. This ignorance forbids analytical forms of the quantities defined in both these equations. In order to get their analytical forms, we have to make additional plausible assumption about the ratio $\displaystyle \left( \frac{v_2}{v_1}\right)$. We assume that $ x $ and $ t $ dependence of $ v_1 $ and $ v_2$ are factorisable and the $ x $ dependent part does not deviate significantly from their values at the initial curve $ t=t_0 $, i.e.
\be
\label{eqn:ch5v2byv1}
\frac{v_2(x,t)}{v_1(x,t)}\approx\frac{v_2(\tau_2)}{v_1(\tau_1)}.f(t),
\ee
where $f(t)$ is some unknown function of $t$.
Using Eq.(\ref{eqn:ch5v2byv1}) in Eqs.(\ref{eqn:ch5Gxt1}) and (\ref{eqn:ch5F2sxt}), we get the gluon momentum distribution and the singlet structure function as :
\begin{eqnarray}
\label{eqn:ch5Gxtsemia}
G(x,t)= G(\tau_1)exp\left[\frac{2 n_f}{7}\left(x^{\left( \frac{t_0}{t}\right)^{9A_f}}-x\right)\right.  & \nonumber \\ \left.-9A_f\left(\frac{11}{12}+\frac{n_f}{18}\right)\ln\frac{t}{t_0}+\ln\frac{1}{x}\left(1-\left(\frac{t_0}{t}\right)^{9A_f}\right)\right]  & \nonumber \\
\times \exp\left[12 A_f \frac{F_2^S(\tau_1)-\frac{3n_f}{14}G(\tau_1)}{G(\tau_1)}\int_{t_0}^t x_1(t')\,ln\frac{1}{x_1(t')} f(t')dt'\right]
\end{eqnarray}
and
\begin{eqnarray}
\label{eqn:ch5F2ssemia}
F_2^S(x,t)=\frac{3 n_f}{14}G(x,t)+\left(F_2^S(\tau_2)-\frac{3 n_f}{14}G(\tau_2)\right)exp\left[9 n_f\left(x-x^{\left(\frac{t_0}{t}\right)^{2A_f}}\right)\right] & & \nonumber \\
\times exp\left[\frac{G(\tau_2)}{F_2^s(\tau_2)-\frac{3 n_f}{14}G(\tau_2)}\int_{t_0}^t c_2'(x_2(t'),t')dt'\right],
\end{eqnarray}
where the function $c_2'(x_2(t'),t')$ appearing on the r.h.s of Eq.(\ref{eqn:ch5F2ssemia}) is given by
\begin{eqnarray}
\label{eqn:ch5c2des}
c_2'(x_2(t'),t') &=&\left( A_f n_f-\frac{27}{49} A_f n_f^2 x_2(t)ln\frac{1}{x_2(t')}+ \frac{27}{14} A_f n_f \left(\frac{11}{12}+\frac{n_f}{18}\right)\right. \nonumber \\
& & \left.-\frac{27}{14} A_f n_f ln\frac{1}{x_2(t')}\right) \frac{1}{t' f(t')} \, .
 \hspace{0.1in}
\end{eqnarray}
The functions $ x_1(t) $ and  $ x_2(t) $ are defined in Eqs.( \ref{eqn:ch5x1t}) and ( \ref{eqn:ch5x2t}). Eq.(\ref{eqn:ch5Gxtsemia}) and Eq.(\ref{eqn:ch5F2ssemia}) are the semi-analytical expressions for the gluon and the singlet structure functions  in terms of the undetermined function $ f(t)$ . For analytical forms, we need to use simple test function for $ f(t)$ which we will discuss in \S\ref{sec:ch5results}. Further, if we expand $ f(t)$ about $ t=t_0$ and retain only the first term, then
\begin{eqnarray}
\label{eqn:ch5ftnot}
f(t)=f(t_0)+(t-t_0)f'(t)\mid_{t=t_0}+ ...
  \approx f(t_0) \, .
\end {eqnarray}
Integration over $ t'$ in Eqs.(\ref{eqn:ch5Gxtsemia}) and (\ref{eqn:ch5F2ssemia}) can now be performed leading to the following analytical forms of gluon and singlet densities valid for $ t $ very close to the initial curve $ t=t_0 $  in terms of one unknown parameter $ f_0=f(t_0)$ :
\begin{eqnarray}
\label{eqn:ch5Gxtana}
G(x,t)=G(\tau_1)\exp\left[\left(\frac{4}{3} \frac{F_2^S(\tau_1)-\frac{3 n_f}{14}G(\tau_1)}{G(\tau_1)}.f_0+\frac{2 n_f}{7}\right) \left(x^{(\frac{t_0}{t})^{9A_f}}-x\right) \right]  & & \nonumber  \\
\times\exp\left[\left(1-(\frac{t_0}{t})^{9 A_f}\right)ln\frac{1}{x}\right]exp\left[-9 A_f\left(\frac{11}{12}+\frac{n_f}{18}\right)ln\frac{t}{t_0}\right]
\end{eqnarray}
and
\begin{eqnarray}
\label{eqn:ch5F2sana}
F_2^S(x,t)& = &\frac{3n_f}{14}G(x,t)+\left(F_2^S(\tau_2)-\frac{3n_f}{14}G(\tau_2)\right)\exp\left[9A_f\left(x-x^{(\frac{t_0}{t})^{2A_f}}\right)\right]  \nonumber \\
 & & \times\exp\left[\frac{G(\tau_2)}{F_2^S(\tau_2)-\frac{3 n_f}{14}G(\tau_2)}\frac{1}{f_0}\left\{A_f n_f\left(1+\frac{27}{14}\left(\frac{11}{12}+\frac{n_f}{18}\right)\right)ln\frac{t}{t_0}\right.\right. \nonumber \\
& & \left. \left.-\frac{27}{98}n_f^2\left(x^{\left(\frac{t_0}{t}\right)^{2A_f}}-x\right)-\frac{27}{28}\left(1-\left(\frac{t_0}{t}\right)^{2A_f}\right)ln\frac{1}{x}\right\}\right].
\end{eqnarray}
For phenomenological test we use our Eqs.(\ref{eqn:ch5Gxtsemia}), (\ref{eqn:ch5F2ssemia}), (\ref{eqn:ch5Gxtana}) and (\ref{eqn:ch5F2sana}).
\subsection{Compatibility with earlier solution \label{secs:ch5compatibility}}
In \cite{PKSPram58} we derived an expression  for the gluon momentum distribution considering only the gluon evolution equation and neglecting the quark singlet part. Therefore, putting $F_2^S = 0$ in Eq.(\ref{eqn:ch5Gxtana}) we should get back the earlier solution. But we find that the earlier solution cannot be exactly recovered from Eq.(\ref{eqn:ch5Gxtana}). The origin of this problem lies in the approximation Eq.(\ref{eqn:ch5pgx}) where we have taken $P_g(x)\approx x\ln\frac{1}{x}$ to be used in Eq.(\ref{eqn:ch5b}). But in our earlier work  we used $P_g(x)$ as $P_g(x)\approx x\left(\ln\frac{1}{x} - \frac{11}{12}\right)$. Using that form of $P_g(x)$ in Eq.(\ref{eqn:ch5b}) and without changing other quantities, the eigenvalues of the matrix $A$ are found to be,
\be
\lambda_1 = \frac{-9 A_f\,x\,\left(-\frac{11}{12} + \ln\frac{1}{x}\right)}{t}
\ee
and 
 \be
\lambda_2 = \frac{-2 A_f\,x\,\ln\frac{1}{x}}{t}
\ee
to be compared with Eq.(\ref{eqn:ch5lambda1}) and Eq.(\ref{eqn:ch5lambda2}). Only the first characteristic curve changes while the second one remains the same.

Following the same procedure as in the derivation of Eq.(\ref{eqn:ch5Gxtana}) with the assumption Eq.(\ref{eqn:ch5v2byv1}) and Eq.(\ref{eqn:ch5ftnot}) we find the the gluon distribution to be
\begin{eqnarray}
\label{eqn:ch5gxt}
G(x,t)& = & G(\tau)\exp\left[-H(x,t)\right] \nonumber \\
& & \times \exp\left[\left(-\frac{11}{12} + \ln\frac{1}{x}\right)\left\{1 - \left(\frac{t_o}{t}\right)^{\frac{12}{\beta_o}}\right\} - \frac{\gamma^2 n_f}{18}\ln\frac{t}{t_o}\right],
\end{eqnarray} 
where the function $H(x,t)$   is
\begin{eqnarray}
\label{eqn:ch5hxt}
H(x,t) &=&\left[ 
\frac{11}{12e^{11/12}}\left\{ \Gamma_1 -\Gamma_2 \right\}  
+\left\{ \exp \left( -\ln \frac{1}{x}\right)\right.\right.\nonumber \\
&& \left.\left. -\exp \left( -\frac{11}{12}+\left( \frac{11}{12}-\ln \frac{1}{x}\right) \left( \frac{t_{0}}{t}\right) ^{\frac{12}{\beta _{0}}}\right) \right\}
\right] \nonumber \\
&&\times \left( \frac{F_2^{S}(\tau )-\frac{3n_{f}}{14}G(\tau )}{G(\tau )}\frac{4}{3}.\,f(t_0)+\frac{2n_{f}}{7}\right)
\end{eqnarray}
and $\tau$ is given by 
\be
\tau = \exp\left[\left(\ln\frac{1}{x} + \frac{11}{12}\right)\left(\frac{t_o}{t}\right)^{9A_f}-\frac{11}{12}\right]
\ee
which is exactly same as we derived in \cite{PKSPram58}.
In Eq.(\ref{eqn:ch5hxt}) $\Gamma_1$ and $\Gamma_2$ are the incomplete Gamma functions
\be
\Gamma_1 = \Gamma\left[0,\; \left(-\frac{11}{12} + \ln\frac{1}{x}\right)\left(\frac{t_o}{t}\right)^{9A_f}\right],
\ee
\be
\Gamma_2  = \Gamma\left[0,\; \left(-\frac{11}{12} + \ln\frac{1}{x}\right)\right]
\ee
which is defined as
\be
\Gamma(0,z) = \int_z^\infty e^{-t}t^{-1}dt
\ee
Putting $F_2^S = 0$ in Eq.(\ref{eqn:ch5hxt}) with the additional assumption that $f\left(t_o\right) = 1$, which is equivalent to the assumption that the ratio $\displaystyle { \frac{v_2}{v_1 } }$ (Eq.\ref{eqn:ch5v1xtbar}) along the characteristic is equal to its value on the initial curve, we find that Eq.(\ref{eqn:ch5gxt}) is reduced to 
\be
\label{eqn:ch5Gxt}
G(x,t) = G(\tau)\left(\frac{1}{x}\right)^{1-\left(\frac{t_o}{t}\right)^{9A_f}}\left(\frac{t_o}{t}\right)^{\frac{A_fn_f}{2}}\exp\left[-\frac{11}{12}\left(1-\left(\frac{t_o}{t}\right)^{9A_f}\right)\right]
\ee
which is exactly the same expression  we obtained earlier in \cite{PKSPram58} without considering the singlet part.
\subsection{Non singlet structure function $ F_2^{NS}(x,t) $ } 
\label{subs:ch5Nonsinglet}
A similar procedure can be followed for the \D for the  non-singlet structure function $ F_2^{NS}$ :
\begin{eqnarray}
\label{eqn:ch5Fns}
\frac{\p F_2^{NS}(x,t)}{\p t}=\frac{A_f}{t}\left[\left\{3+4ln(1-x)\right\}F_2^{NS}(x,t)\right. & & \nonumber \\
 \left.
 +2\int_x^1\frac{dz}{1-z}\left\{(1+z^2)F_2^{NS}\left(\frac{x}{z},t\right)-2F_2^{NS}(x,t)\right\}\right].
\end{eqnarray}
Following the approximation similar to Eqs.(\ref{eqn:ch5F2seriesApp}) and (\ref{eqn:ch5GseriesApp}) we get,
\be
\label{eqn:ch5Fnsapp}
F_2^{NS}(\frac{x}{z},t)\approx F_2^{NS}(x,t)+\left(x\sum_{k=1}^\infty u^k\right)\frac{\p F_2^{NS}(x,t)}{\p x}.
\ee
Now putting Eq.(\ref{eqn:ch5Fnsapp}) in Eq.(\ref{eqn:ch5Fns}) we get
\begin{eqnarray}
\label{eqn:ch5Fnsexp}
\frac{\p F_2^{NS}(x,t)}{\p t}=\frac{A_f}{t}\left[\left\{3+4ln(1-x)\right\}F_2^{NS}(x,t)+2\int_x^1\frac{dz}{1-z}(z^2-1)F_2^{NS}(x,t) \right. & & \nonumber \\
\left.
+2\int_x^1\frac{dz}{1-z}(1+z^2)(x\sum_{k=1}^\infty u^k) \frac{\p F_2^{NS}(x,t)}{\p x}   \right].
\end{eqnarray}
Carrying out the integration in $ z$ and neglecting terms $ O(x^2)$ and higher, we can write Eq.(\ref{eqn:ch5Fnsexp}) as 
\be
\label{eqn:ch5Fnspde}
\frac{\p F_2^{NS}(x,t)}{\p t}-\frac{8A_f}{3}\frac{x}{t} \frac{\p F_2^{NS}(x,t)}{\p x}=\frac{A_f \left\{4ln(1-x)+2x\right\}}{t}F_2^{NS}(x,t)
\ee
Eq.(\ref{eqn:ch5Fnspde}) is a partial differential equation for the non-singlet function $F_2^{NS}(x,t) $ with respect to the variables $x$ and $t$. The characteristic curve of the Eq.( \ref{eqn:ch5Fnspde}) is given by the solution of the differential equation
\be
\label{eqn:ch5Fnscheq}
\frac{d x_3(t)}{d t}=-\frac{8A_f}{3}\frac{x}{t}
\ee
Assuming that the characteristic curve passes through a point $ ( \tilde{x},\tilde{t})$ i.e $ x_3(\tilde{t})=\tilde{x}$ in the $x - t$ space, we get the solution of Eq.(\ref{eqn:ch5Fnscheq}) to be
\be
\label{eqn:ch5Fnschsol}
ln\frac{x_3(t)}{\tilde{x}}=-\frac{8A_f}{3}ln\frac{t}{\tilde{t}}\; .
\ee
If the characteristic curve cuts the initial curve $ t=t_0 $ at a point $ x_3(t_0)=\tau_3 $, then Eq.(\ref{eqn:ch5Fnschsol}) gives
\be
ln\frac{\tau_3}{\tilde{x}}=-\frac{8A_f}{3}ln\frac{t_0}{\tilde{t}} \, ,
\ee
so that
\be
\label{eqn:ch5tau3}
\tau_3=\tilde{x}\left(\frac{t_0}{\tilde{t}}\right)^{-\frac{8A_f}{3}}.
\ee
Dropping the `hats'  over $ x $ and $ t $, the equation of the characteristic is
\be
\label{eqn:ch5cheq3}
x_3(t)=\tau_3\left(\frac{t_0}{t}\right)^{\frac{8A_f}{3}}.
\ee

On using Eq.(\ref{eqn:ch5Fnscheq}) in Eq.(\ref{eqn:ch5Fnspde}), the left hand side becomes an ordinary derivative with respect to $ t$ and the equation becomes an ordinary differential equation:
\be
\label{eqn:ch5Fnsode}
\frac{d F_2^{NS}(x(t),t)}{d t}=c^{NS}(x(t),t)F_2^{NS}(x(t),t) \, ,
\ee
where
\be
\label{eqn:ch5cns}
c^{NS}(x(t),t)=\frac{A_f\{4 ln(1-x(t))+2x(t)\}}{t}.
\ee
Integrating Eq.(\ref{eqn:ch5Fnsode}) over $ t $ from $ t_0$ to $ \bar{t}$ along the characteristic curve $ x_3(t)$ (Eq \ref{eqn:ch5cheq3}) and finally dropping the bars over $ x $ and $ t $ we get the solution for the non-singlet as :
\begin{eqnarray}
\label{eqn:ch5Fnsana}
F_2^{NS}(x,t)& = & F_2^{NS}(\tau_3)\exp\left[\frac{3}{4A_f} x \left(\left(\frac{t}{t_0}\right)^{\frac{8A_f}{3}}-1\right)\right. \nonumber \\
& & \left.-\frac{3}{2A_f}\left\{\sum_{k=1}^{\infty}\frac{x^k}{k^2}\left(\left(\frac{t}{t_0}\right)^{\frac{8A_fk}{3}}-1\right)\right\}\right].
\end{eqnarray}
Eqs.(\ref{eqn:ch5Gxtana}), (\ref{eqn:ch5F2sana}) and (\ref{eqn:ch5Fnsana}) are the analytical solutions of the  \D within the present formalism. While Eqs.(\ref{eqn:ch5Gxtana}) and (\ref{eqn:ch5F2sana}) contain a free parameter $ f_0 $, Eq.(\ref{eqn:ch5Fnsana}) is a parameter free solution.

Using our results derived in this section, we will calculate the proton structure function $ F_2^p $ which is related to the singlet and the non-singlet structure function by the equation
\be
\label{eqn:ch5F2p}
F_2^p=\frac{3}{18}F_2^{NS}+\frac{5}{18}F_2^S\,.
\ee
We discuss in the next section the phenomenological consequences of our results derived in this section. 
\begin{figure}
\vspace{-0.3in}
\begin{center}
\includegraphics[width=3.5in]{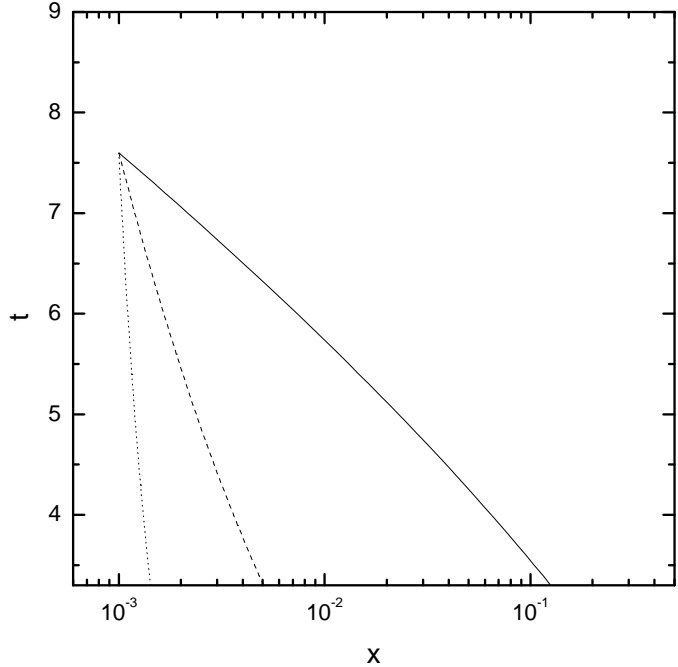}
\end{center}
\caption[Characteristic curves given by Eq.(\ref{eqn:ch5x1t}),Eq.(\ref{eqn:ch5x2t}) and Eq.(\ref{eqn:ch5cheq3}) from an initial line $t=t_0(=ln\frac{Q_0^2}{\Lambda^2})$ to the point $\bar{x}=0.001,\bar{t}=7.65 (Q^2=80GeV^2)$ ]{Characteristic curves given by Eq.(\ref{eqn:ch5x1t})(solid line),Eq.(\ref{eqn:ch5x2t})(dotted line) and Eq.(\ref{eqn:ch5cheq3})(dash-dotted line) drawn from the initial line $t=t_0(=ln\frac{Q_0^2}{\Lambda^2})$ to the point $\bar{x}=0.001,\bar{t}=7.65 (Q^2=80GeV^2)$. Here we take $n_f=4,A_f=\frac{4}{25}, Q_0^2=1 GeV^2$ and $\Lambda^2=0.200 GeV^2$.}
\label{fig:ch5fig1}
\end{figure}
 \begin{figure}
\begin{center}
\includegraphics[width=6.2in,height=6.0in]{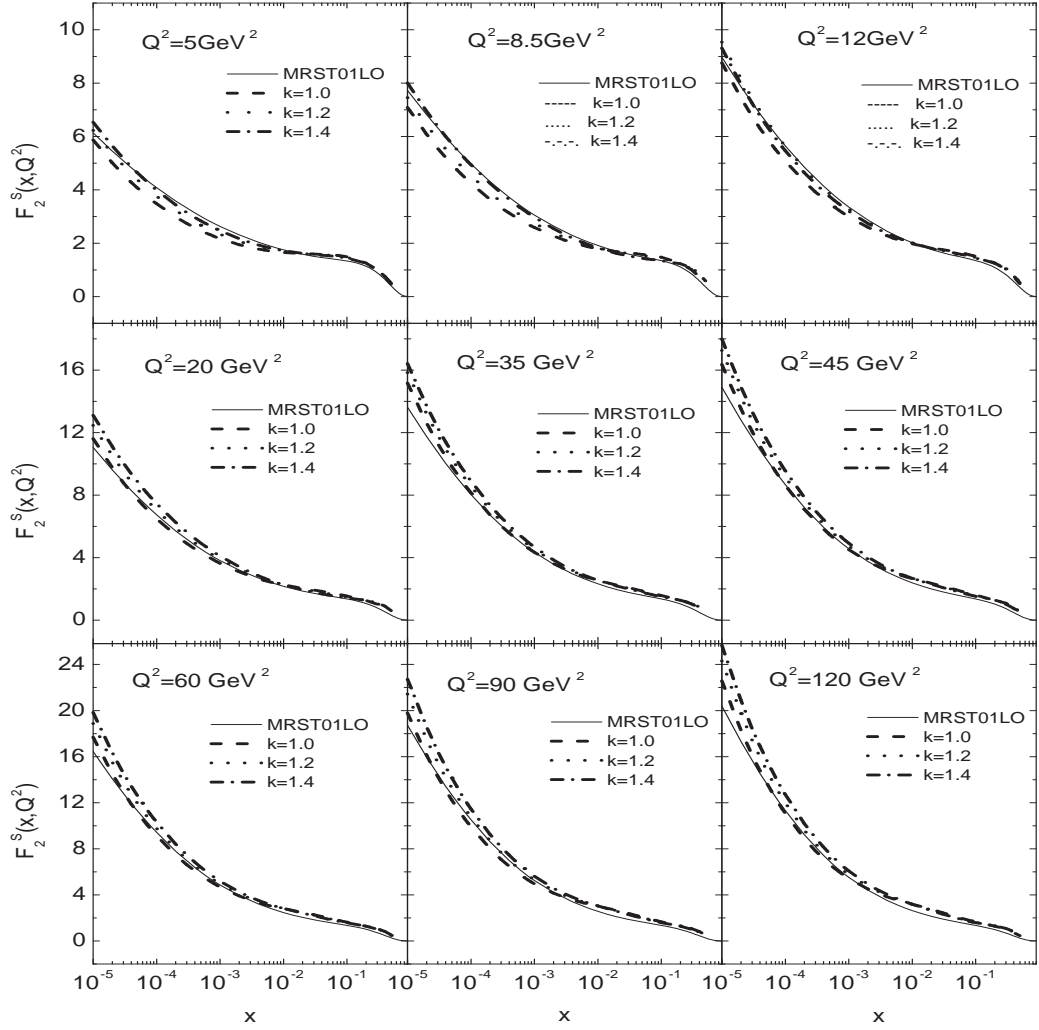}
\end{center}
\caption[Semi-analytical singlet structure function $F_2^S(x,t)$ (Eq.\ref{eqn:ch5F2ssemia}) compared with the exact MRST01LO solutions at nine  different $Q^2$ ]{Semi-analytical singlet structure function $F_2^S(x,t)$ (Eq.\ref{eqn:ch5F2ssemia}) compared with the exact MRST01LO\cite{MRST2001} solutions (solid line) at nine  different $Q^2$ for three different values of the parameter $ k=1.0$ (dashed line ), 1.2 (dotted line )  and  1.4 (dash-dotted line.)}
\label{fig:ch5fig2}
\end{figure}
\begin{figure}
\begin{center}
\includegraphics[width=5.9in]{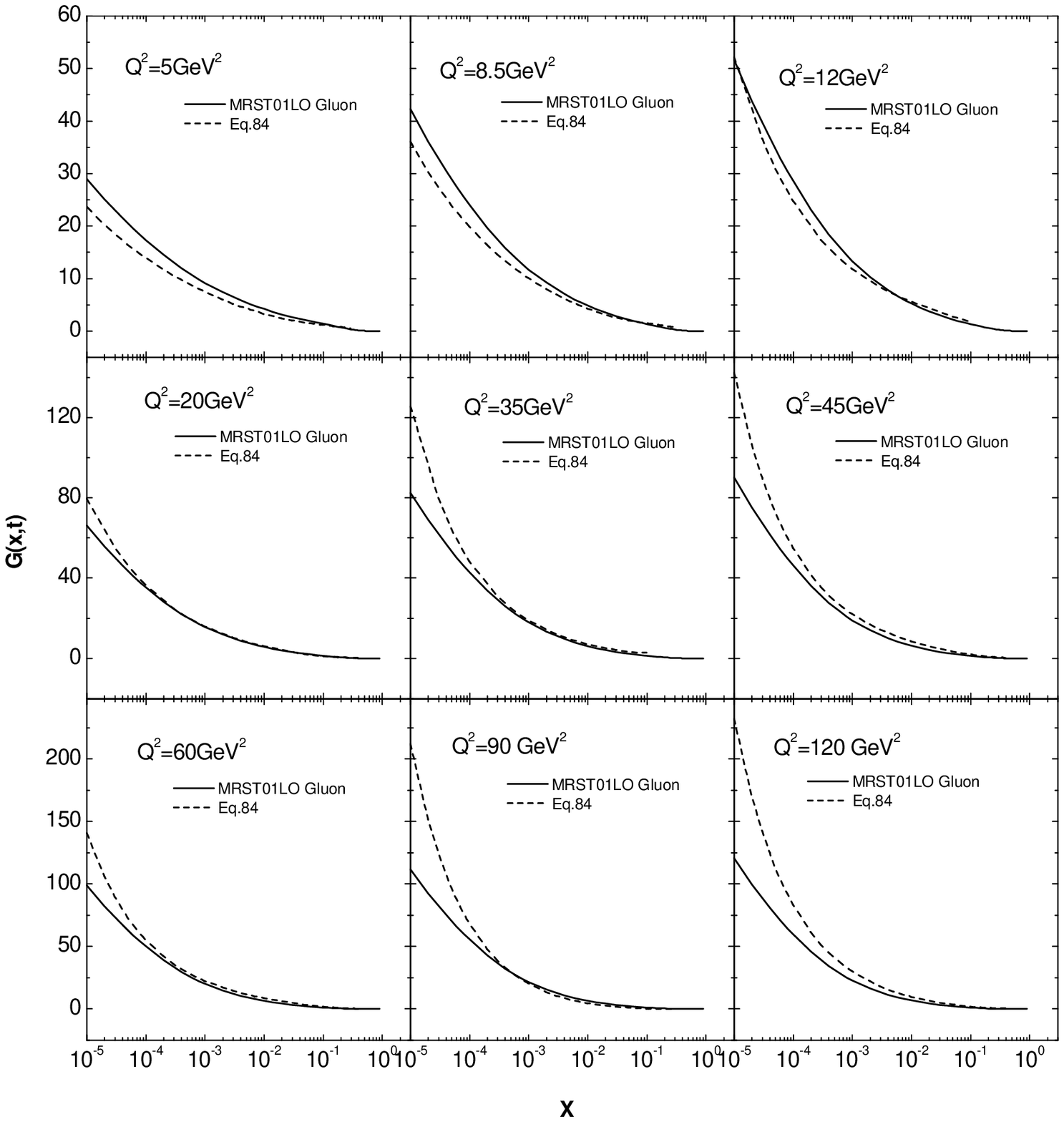}
\end{center}
\caption[Semi-analytical Gluon momentum distribution given by Eq.(\ref{eqn:ch5Gxtsemia}) compared with the exact  MRST01LO solutions at nine different  $Q^2$ ]{Semi-analytical Gluon momentum distribution given by Eq.(\ref{eqn:ch5Gxtsemia}) compared with the exact  MRST01LO\cite{MRST2001} solutions at nine different representative $Q^2$. The gluon (Eq.\ref{eqn:ch5Gxtsemia}) is shown for only one value of $k=1.2$ as discussed in the text.}
\label{fig:ch5fig3}
\end{figure}
\section{Results and discussion \label{results}}
\label{sec:ch5results}

In  \S\ref{subs:ch5solbymethodofch} we have derived semi-analytical [Eq.(\ref{eqn:ch5Gxtsemia}) and Eq.(\ref{eqn:ch5F2ssemia})] and analytical [Eq.(\ref{eqn:ch5Gxtana}) and Eq.(\ref{eqn:ch5F2sana})] solutions of the gluon and the singlet and in  \S\ref{subs:ch5Nonsinglet} obtained an analytical solution [Eq.(\ref{eqn:ch5Fnsana})] for non-singlet structure function $F_2^{NS}$ in LO.Let us now compare them with the exact solutions of the DGLAP equations to see in what region of $x-Q^2$ space they agree with the exact ones within a few (say 15\%) percent. We choose  MRST2001 LO \cite{MRST2001} solutions for comparison. 
For evolution of Eq.(\ref{eqn:ch5Gxtsemia}), Eq.(\ref{eqn:ch5F2ssemia}),  Eq.(\ref{eqn:ch5Gxtana}), Eq.(\ref{eqn:ch5F2sana})  and   Eq. (\ref{eqn:ch5Fnsana}) we take the inputs from the above ref\cite{MRST2001}. We note that in our expressions [Eq.(\ref{eqn:ch5Gxtsemia}), Eq.(\ref{eqn:ch5F2ssemia}), Eq.(\ref{eqn:ch5Gxtana}), Eq.(\ref{eqn:ch5F2sana}) and Eq.(\ref{eqn:ch5Fnsana})] we need to take the inputs as functions of $\tau_i$ (i=1,2,3) rather than of $x$, where $\tau_i$ are the points at  which the characteristic curves defined in Eqs.(\ref{eqn:ch5x1t}), (\ref{eqn:ch5x2t}) and (\ref{eqn:ch5Gxt}) cut the initial line $t =t_o$ ($t=\ln\frac{Q^2}{\Lambda^2}$) respectively. For $n_f=4$ and $A_f= \frac{4}{25}$ $\left(A_f=\frac{4}{3\beta_o}\right)$, these are $\tau_1=x^{\left(\frac{t_o}{t}\right)^{\frac{36}{5}}}$, $\tau_2=x^{\left(\frac{t_o}{t}\right)^{\frac{8}{25}}}$ and $\tau_3=x^{\left(\frac{t_o}{t}\right)^{\frac{32}{75}}}$ respectively. The inputs thus acquire a $t$ dependence also in addition to $x$ and are obtained by a formal replacement $x\rightarrow \tau_i$ in the MRST2001 LO \cite{MRST2001} input distributions.

To evaluate numerically the semi-analytical expressions Eq.(\ref{eqn:ch5Gxtsemia}) and Eq.(\ref{eqn:ch5F2ssemia}), we need the unknown function $f(t)$ defined in Eq.(\ref{eqn:ch5v2byv1}). As $f(t)$ is the ratio of the $t$ dependent part of $v_2$ and $v_1$ defined in Eqs.(\ref{eqn:ch5v1xt}) and (\ref{eqn:ch5v2xt})  we assume it to be of the form $f(t)=\left(\ln t\right)^k$, where $k$ is a parameter to be determined by comparison with the exact solutions. Varying $k$ continuously from negative to positive values, we find that only in a limited range of positive values from $k\simeq 1.0$ to $k\simeq 1.4$, the semi-analytical solutions $G(x,t)$ [Eq.(\ref{eqn:ch5Gxtsemia})] and $F_2^S(x,t)$ [Eq.(\ref{eqn:ch5F2ssemia})] compare well with the exact ones. In Fig.2 we show  $F_2^S(x,t)$ [Eq.(\ref{eqn:ch5F2ssemia})] as function of $x$ at nine different representative $Q^2$ for three different values of $k=1.0,1.2 $ and $1.4$.
 We see that the value of $k$ is crucial for the consistency of our solution {\it vis-a-vis} the exact ones and different values of $k$ seem to suit at different $Q^2$. Our prediction remains below the exact solution for $Q^2$ up to $\approx 20 GeV^2$ and $x\leq 10^{-2}$. But above $Q^2\geq 60 GeV^2$ and $x\leq 10^{-2}$, it overshoots the exact solution and rises faster with decreasing $x$.  However, above $x\geq 10^{-2}$, our prediction lies above the exact ones for all the $Q^2$ values shown.Taking the graph for $k=1.2$, we calculate the percentage deviation of our solution from the exact one at seven different equally spaced $x$ at each $Q^2$ and find that in the range $6.5\, GeV^2\leq Q^2\leq 60\, GeV^2 $ and $ 10^{-4}\leq x \leq 10^{-2}$,  our solution (Eq.(\ref{eqn:ch5F2ssemia})) is within 15\% of the exact one.

 Taking $k=1.2$ we evolve the  gluon given by Eq.(\ref{eqn:ch5Gxtsemia}) and compare with the exact MRST01 \cite{MRST2001} solution in Fig.3 at nine different $Q^2$. We see almost similar results as in the case of $F_2^S$. Below $Q^2\simeq 20\, GeV^2$ the predicted gluon density remains lower than the exact one in the small $x$ $\left( x\leq 10^{-1}\right)$ region but as $Q^2\geq 20 GeV^2$, it begins to overshoot the exact solution, so that at $Q^2\simeq 60 GeV^2$ and $x\simeq 10^{-4}$, the percentage deviation from the exact one becomes more than $15\%$. For $x\geq 10^{-4}$ and $5\, GeV^2\leq Q^2\leq 60\, GeV^2 $, the predicted gluon(Eq.\ref{eqn:ch5Gxtsemia}) is within 15\% of the exact solution.

\begin{figure}
\begin{center}
\includegraphics[width=5.8in]{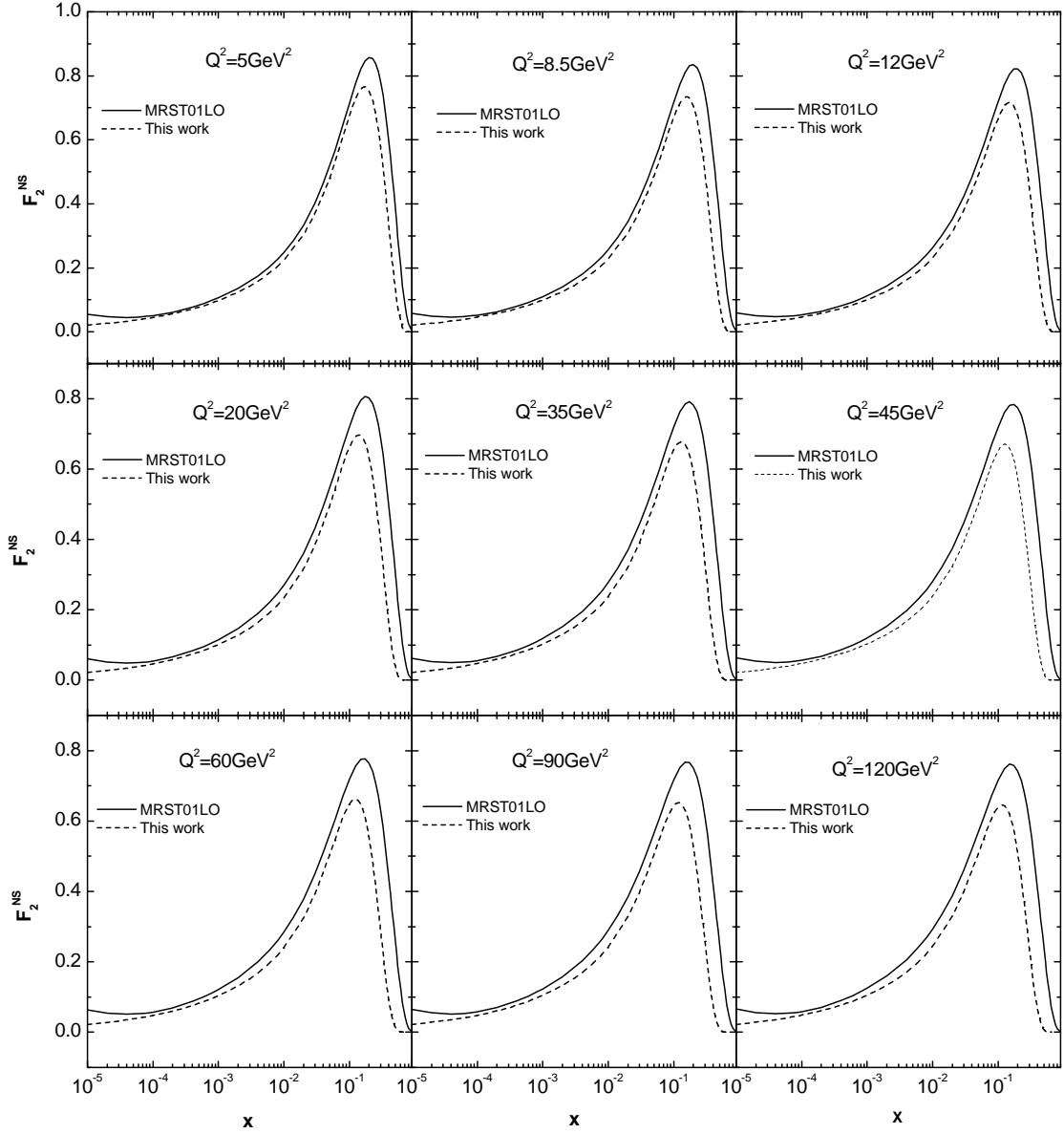}
\end{center}
\caption[Non-singlet structure function $F_2^{NS}$ (Eq.\ref{eqn:ch5Fnsana}) compared with the exact solution ]{Non-singlet structure function $F_2^{NS}$ (Eq.\ref{eqn:ch5Fnsana}) compared with the exact solution \cite{MRST2001} at the same values of $Q^2$ as in Fig.(\ref{fig:ch5fig2}) and Fig.(\ref{fig:ch5fig3}).}
\label{fig:ch5fig4}
\end{figure}

\begin{figure}
\begin{center}
\includegraphics[width=5.6in]{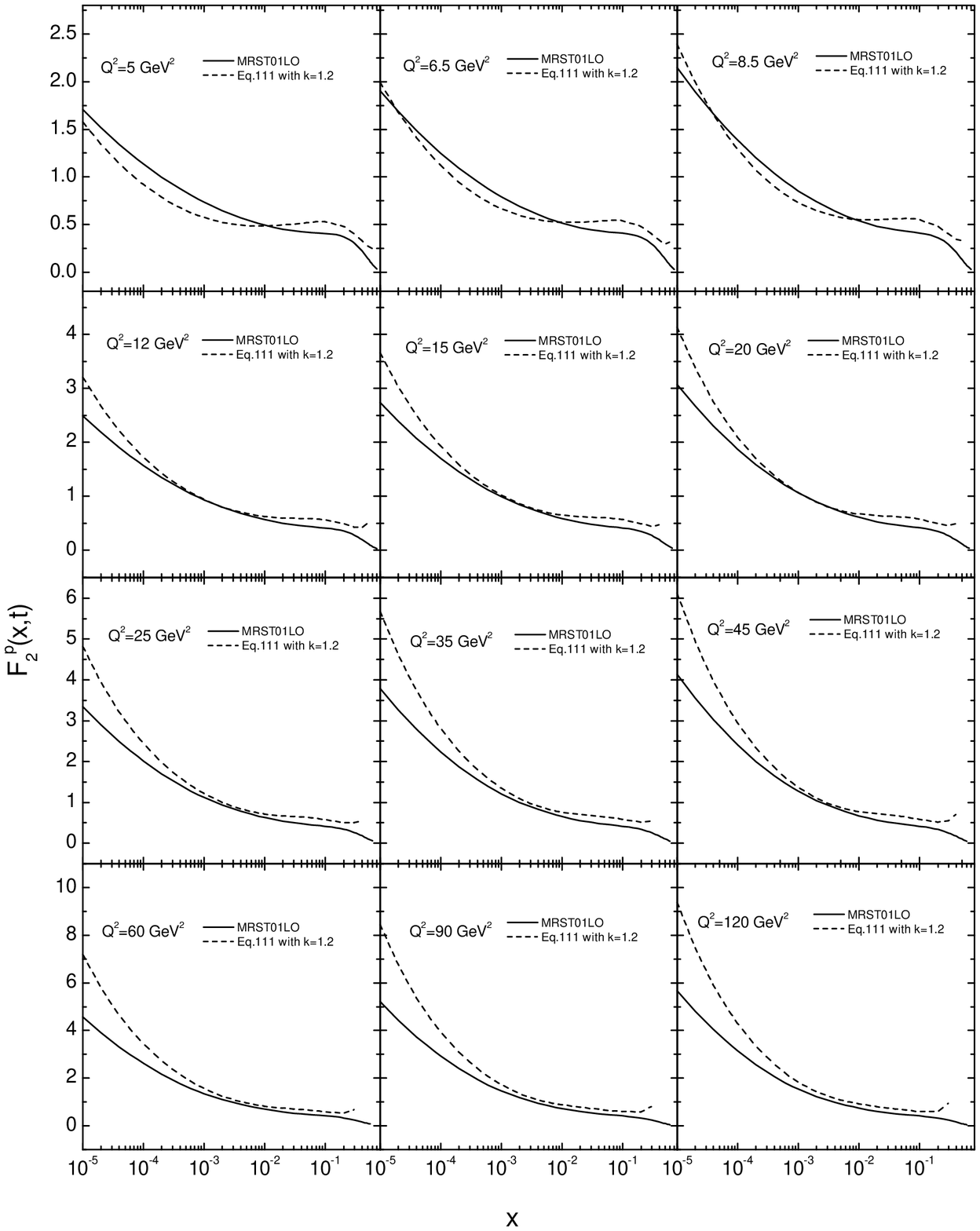}
\end{center}
\vspace{-0.10in}
\caption[Proton structure function $F_2^p(x,t)$ calculated using Eq.(\ref{eqn:ch5F2ssemia}) and Eq.(\ref{eqn:ch5Fnsana}) in Eq.(\ref{eqn:ch5F2p}) compared with the exact MRST01LO solution ]{Proton structure function $F_2^p(x,t)$ calculated using Eq.(\ref{eqn:ch5F2ssemia}) and Eq.(\ref{eqn:ch5Fnsana}) in Eq.(\ref{eqn:ch5F2p}) compared with the exact MRST01LO\cite{MRST2001} solution.Only one value of the free parameter $k (=1.2)$ is used as discussed in the text.}
\label{fig:ch5fig5}
\end{figure}

 In Fig.4 the non-singlet solution $F_2^{NS}(x,t)$ (Eq.(\ref{eqn:ch5Fnsana})) is shown along with the exact MRST2001 LO solution. Unlike the singlet and the gluon, the non-singlet solution  is free from any free parameter. The qualitative features of Eq.(\ref{eqn:ch5Fnsana}) and the exact solution are identical, the exact one remaining always slightly higher. 

Combining  Eq.(\ref{eqn:ch5F2ssemia}) and Eq.(\ref{eqn:ch5Fnsana}) in Eq.(\ref{eqn:ch5F2p}) we  calculate the proton structure function $F_2^p$. The result is compared with the exact solution \cite{MRST2001} in Fig.5 at twelve different representative values of $Q^2$ as a function of $x$. Taking some (eight) equally spaced $x$ values at each $Q^2$, we calculate the percentage deviation of our result from the exact one and find that for $5\, GeV^2\leq Q^2\leq 45\, GeV^2 $ and $ 10^{-4}\leq x \leq 10^{-2}$, our $F_2^p$ is within 15\% of the exact MRST01LO solutions.

Next we study the analytical solutions Eq.(\ref{eqn:ch5Gxtana}) and Eq.(\ref{eqn:ch5F2sana}) which contain a free parameter $f_0$ and are expected to be valid at $t$ very close to $t_0$ $\left(\equiv \ln \frac{Q_0^2}{\Lambda^2}\right)$. Here also we find  the value of the free parameter $f_0$ such that our solutions are compatible with the exact ones in a definite $ x- Q^2 $ range.We find that for $ f_0=0.8$, our approximate analytical solutions reproduce the exact results in a limited range of $x (10^{-4} \leq x  \leq 10^{-2})$ and $Q^2$($5GeV^2 \leq Q^2 \leq 60 GeV^2 )$ within $15 \%$.

\section{Conclusion}
\label{con:ch5conclusion}
We have presented in this paper solutions of the singlet structure function $F_2^S(x,t)$ and the gluon momentum distribution $G(x,t)$ valid to be at small $x$. Applying the method of characteristics , the LO coupled DGLAP evolution equations are solved in the $x-t$ space. The results are presented both in analytic and semi-analytic forms and compared with the exact MRST2001LO \cite{MRST2001} solutions to find the range of validity of the solution. Applying the same method, LO evolution equation for the non-singlet structure function $F_2^{NS}(x,t)$ is also solved analytically.

In order to obtain the solutions for $F_2^S$ and $G$, we however need additional information about the ratio $\displaystyle{\frac{v_2}{v_1}=\frac{F_2^S(x,t)}{G(x,t)}-\frac{3 n_f}{14}}$.This is achieved through an unknown function $f(t)$ (Eq.\ref{eqn:ch5v2byv1}) or an unknown parameter $f_0$(Eq.\ref{eqn:ch5ftnot}). Through their suitable choices, $f(t)=(lnt)^{1.2} $ or $f_0=0.8 $, we found regions of $x-Q^2$ where our approximate solutions are compatible with the exact ones within 15 \%.While this additional information needed is an inherent limitation of the present formalism, still it can be considered an improvement over our previous work in refs \cite{JSDKCGMPLB403,DKCASPram33,DKCJKSPram38} where identical $t-$ evolution were assumed for gluon and singlet distributions unlike Eq.(\ref{eqn:ch5v2byv1}).

In this paper  we have not compared our results with data directly, since exact LO solutions like \cite{MRST2001} reproduce the data correctly.A similar agreement with data in the entire $x-Q^2 $ range of HERA  of  our approximate solution presumably need a new fit of the input distributions and other free parameters/functions like $f(t)$ and $f_0$. The same inputs as used \cite{MRST2001} by us to compare our results with exact ones are not sufficient to reproduce the entire data, as is evident from our analysis.This is understandable, since approximate solutions refer to a different procedure with different evolution and thus to a possibly different input.Such a possibility is currently under investigation.

\end{document}